\documentclass[aps, prx, twocolumn,longbibliography,floatfix,superscriptaddress]{revtex4-1}

\usepackage{graphicx}
\usepackage{dcolumn}
\usepackage{bm}
\usepackage{amsmath,amssymb,amstext,booktabs}

\begin{document}


\title{Measuring the thermodynamic cost of timekeeping}

\author{A.N. Pearson}%
\thanks{These authors contributed equally to this work}
\affiliation{Department of Materials, University of Oxford, Parks Road, Oxford OX1 3PH, United Kingdom}%
 
\author{Y. Guryanova}%
   \thanks{These authors contributed equally to this work}
\affiliation{%
  Institute for Quantum Optics and Quantum Information (IQOQI), Austrian Academy of Sciences, A-1090 Vienna, Austria}%

\author{P. Erker}%
\thanks{These authors contributed equally to this work}
\affiliation{%
  Institute for Quantum Optics and Quantum Information (IQOQI), Austrian Academy of Sciences, A-1090 Vienna, Austria}%

\author{E.A. Laird}  
\affiliation{%
Department of Physics, Lancaster University, Lancaster, LA1 4YB, United Kingdom}%

\author{G.A.D. Briggs}%
\affiliation{%
 Department of Materials, University of Oxford, Parks Road, Oxford OX1 3PH, United Kingdom}%

\author{M. Huber}%
\email{marcus.huber@univie.ac.at}
\affiliation{%
  Institute for Quantum Optics and Quantum Information (IQOQI), Austrian Academy of Sciences, A-1090 Vienna, Austria}%
\author{N. Ares}%
\email{natalia.ares@materials.ox.ac.uk}
\affiliation{%
 Department of Materials, University of Oxford, Parks Road, Oxford OX1 3PH, United Kingdom}%

\date{\today}

\begin{abstract}

All clocks, in some form or another, use the evolution of nature towards higher entropy states to quantify the passage of time. Due to the statistical nature of the second law and corresponding entropy flows, fluctuations fundamentally limit the performance of any clock. This suggests a deep relation between the increase in entropy and the quality of clock ticks. Indeed, minimal models for autonomous clocks in the quantum realm revealed that a linear relation can be derived, where for a limited regime every bit of entropy linearly increases the accuracy of quantum clocks. But can such a linear relation persist as we move towards a more classical system? 
We answer this in the affirmative by presenting the first experimental investigation of this thermodynamic relation in a nanoscale clock. We stochastically drive a nanometer-thick membrane and read out its displacement with a radio-frequency cavity, allowing us to identify the ticks of a clock.
We show theoretically that the maximum possible accuracy for this classical clock is proportional to the entropy created per tick, similar to the known limit for a weakly coupled quantum clock but with a different proportionality constant. We measure both the accuracy and the entropy. Once non-thermal noise is accounted for, we find that there is a linear relation between accuracy and entropy and that the clock operates within an order of magnitude of the theoretical bound.

\end{abstract}

 \newcommand{\qbar}{\overline{q}}
\newcommand{\wwidth}{183mm}

\newcommand{\Th}{\ensuremath{T_\mathrm{h}} }
\newcommand{\Tc}{\ensuremath{T_\mathrm{c}} }
\newcommand{\TN}{\ensuremath{T_\mathrm{N}} }
\newcommand{\Qf}{\ensuremath{Q_\mathrm{f}} }
\newcommand{\ff}{\ensuremath{f_0} }
\newcommand{\RR}{\ensuremath{R_0} }

\newcommand{\VS}{\ensuremath{V_\mathrm{S}} }
\newcommand{\ttick}{\ensuremath{t_\mathrm{tick}} }
\newcommand{\Stick}{\ensuremath{S_\mathrm{tick}} }

\newcommand{\Vin}{\ensuremath{V_\mathrm{in}} }
\newcommand{\Vout}{\ensuremath{V_\mathrm{out}} }
\newcommand{\VWN}{V_\text{WN}}
\newcommand{\Pc}{\ensuremath{P_\mathrm{c}} }
\newcommand{\Vc}{\ensuremath{V_\mathrm{c}} }
\newcommand{\fc}{\ensuremath{f_\mathrm{c}} }
\newcommand{\fm}{\ensuremath{f_\mathrm{m}} }
\newcommand{\fP}{\ensuremath{f_\mathrm{P}} }
\newcommand{\fE}{\ensuremath{f_\mathrm{E}} }
\newcommand{\PSB}{\ensuremath{P_\mathrm{SB}} }
\newcommand{\trr}{\ensuremath{t_\mathrm{r}} }
\newcommand{\SVVN}{\mathcal{S}_{VV}^\mathrm{(N)}}
\newcommand{\SVV}{\mathcal{S}_{VV}}

\newcommand{\kB}{\ensuremath{k_\mathrm{B}} }                           

\newcommand{\NC}{\ensuremath{N_\mathrm{C}} }
\newcommand{\NJ}{\ensuremath{N_\mathrm{J}} }
\newcommand{\NM}{\ensuremath{N_\mathrm{M}} }

\newcommand{\CD}{\ensuremath{C_\mathrm{D}} }
\newcommand{\CM}{\ensuremath{C_\mathrm{M}} }
\newcommand{\CC}{\ensuremath{C_\mathrm{C}} }
                            
\maketitle

\section{Introduction}
\label{sec:introduction}

By modern standards, the accuracy with which we can keep time is truly astonishing; nowadays the best atomic clocks keep time to an accuracy of approximately one second in every one-hundred million years \cite{Nicholson2015}. This is more accurate than any physical constant we have ever measured (for example, the magnetic moment of an electron $g$ is known to 12 digits \cite{Odom2006}), and better than computer arithmetic which has an accuracy of 16 digits for 64-bit calculations \cite{Rajaraman2016}. 
Atomic clocks run by the rules of quantum mechanics, targeting a specific hyperfine transition in an atom's energy spectrum; yet despite the great progress in keeping time, surprisingly little is known about the relation between quantum clocks and thermodynamics. Famously invariant under time-reversal, the equations of quantum mechanics provide little explanation for the passage of time, whereas the theory of thermodynamics, although elucidating little more on the same front, does at least leave some entropic signatures \cite{Erksi,Erkerthesis,Barato_2016,milburn2017quantum}.
One of the milestones at the intersection of the two fields is to derive a quantitative relation between the second law of thermodynamics and the flow of time. Investigations in this direction are also a vital component in our understanding of quantum thermodynamics, a field focused on the investigation, analysis and design of machines on the quantum scale, to which clocks are no exception~\cite{Goold}. 

Alongside philosophical and conceptual curiosities, clocks constitute an intrinsic component in the operation of numerous systems, from the clocks used to time the gates on a desktop CPU to the clocks necessary for determining your GPS coordinates. 
In the quantum regime, as opposed to the classical case, the  thermodynamic cost associated with the precise control of a system is comparable to the energy scale of the system itself \cite{Manzano_2019,Abah_2019,Clivaz_2019}. 
For example, the cycles of a quantum Otto engine need to be controlled by a microscopic autonomous clock~\cite{shorty,ralphy}; a device that produces a stream of ticks without any timing input or external control. The energetic cost of running this clock is comparable to the energetic output of the engine and thus can no longer be neglected.
These clocks have been studied rigorously from the perspective of open quantum systems, where it has been shown that their performance with respect to the resources they consume is subject to particular relations, as well as trade-offs \cite{Erksi}.
One of the challenges in deriving such relations from microscopic thermodynamic principles is that reasonably large systems are required for irreversible dynamics to emerge~\cite{Gogolreview1,Gogolreview2}. In developing relations and trade-off models, we are forced to make  
assumptions about the underlying parameters and system dynamics~\cite{Mitchy,Goold}. At the other end of the scale, in the classical domain, it is difficult to keep track of thermodynamic costs because the systems become large and complex. 

\begin{figure*}
\includegraphics[scale=0.4]{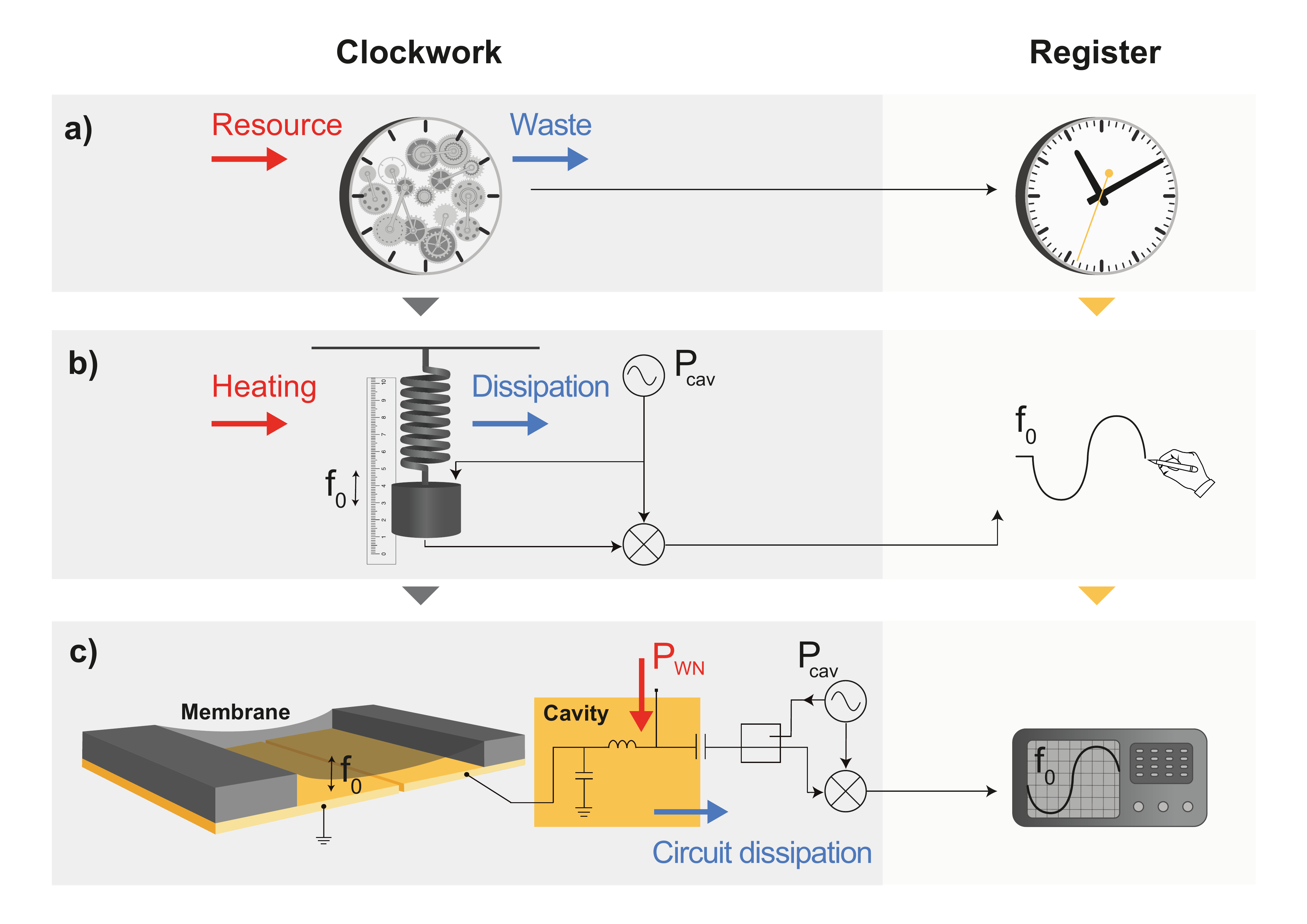}
\caption{\label{fig:1} a) For timekeeping, the clockwork consumes resources, part of which are lost as waste. The hands of the clock register the clock's ticks. b) Simple mechanical clock. Here, a mass is suspended from a spring and the heat from the environment excites the mass' motion at frequency $f_0$. These vibrations are probed by a signal of power $P_\mathrm{cav}$. This system (the clockwork) generates a periodic signal, which is registered to identify the clock's ticks. c) A schematic of our electromechanical system acting as a clock. A nm-thick membrane is driven by a white noise signal of power $P_\mathrm{WN}$. The membrane's vibrations are probed by an RF cavity driven with a signal of power $P_\mathrm{cav}$. The cavity output signal, and thus the clock tick's, are registered by an oscilloscope.}
\end{figure*}

In this article we experimentally explore the thermodynamic costs of timekeeping by directly measuring both the accuracy and the entropy generation associated with a simple nano-electromechanical clock.
This system allows us to investigate the relation between the resources supplied to the clock, in the form of work and heat, and the corresponding accuracy.

A clock, like any thermodynamic machine, operates by consuming a resource and creating waste in the form of entropy (Fig.~\ref{fig:1}(a)).
Its useful output is a train of ticks which can be counted by a register.
Previous theoretical work, based on particular models of classical~\cite{Barato2015} and quantum~\cite{Erksi} clocks, has predicted that in those particular models there is a fundamental price to timekeeping: the more regular and frequent the ticks, the greater the rate at which the clock must create entropy.

This work experimentally and theoretically studies a new kind of classical clock which realises this thermodynamic process.
The clock is based on a simple optomechanical model (Fig.~\ref{fig:1}(b)), in which the Brownian motion of a mechanical resonator is monitored using an electronic cavity interferometer.
Each mechanical oscillation identified by the interferometer corresponds to one tick.
The clock is driven by the work performed to illuminate the cavity and by the heat transferred from the hot resonator to the cold measurement electronics.
While the accuracy can be improved by increasing either the mechanical amplitude or the electrical illumination power, in both cases this leads to greater heat dissipation and therefore increased entropy, as explained in Section~\ref{sec:theory} and  Appendix~\ref{App:ClassicalModel}.

This model is realised as shown in Fig.~\ref{fig:1}(c).
The mechanical resonator is a high-quality silicon nitride membrane vibrating in its fundamental flexural mode.
To excite quasi-Brownian motion, the membrane is driven by a white-noise electrical signal, which acts as an effective thermal bath that raises the mechanical mode temperature~\footnote{White noise is a high-entropy signal and, while it is not straightforward to assign a temperature to it, we don't require a low entropy source of work to prepare it (one can expect that it is reasonably abundant in out-of-equilibrium environments). The notion of temperature itself often becomes ill defined in microscopic contexts, as deviations from thermal equilibrium are more frequent and noticeable and selective coupling to certain frequencies can yield multiple notions of temperature even for an ideal black body.}.
To monitor the membrane's displacement, it is capacitively coupled to a radio-frequency (RF) cavity operated in an optomechanical readout circuit~\cite{Lehnert2014,Bagci2014, Ares2016b, Brown2007,Faust2012,Pearson2020}.
The voltage output of this circuit is proportional to the instantaneous displacement.
This output is recorded using an oscilloscope which acts as the clock register.
Each completed oscillation, identified by an upward zero-crossing of the voltage record, represents one tick of the clock.


We used our setup to test the relation between the resources used to power the clock and its accuracy. The accuracy was determined by an algorithm which marked the instance at which a tick (a particular behavioural signature of the membrane's motion) occurred. We then looked at the accuracy of the optomechanical system for a range of white noise driving power and compared it to the prediction of a classical clock model. In order to make this comparison, we associated the system's resources to the clock's total entropy production. Our results confirm clear proportionality between the driving power (the resource) and the periodicity of the cavity output signal (the accuracy), which is the trademark response predicted by both a quantum and a classical clock model. This finding suggests that fundamental relations for the thermodynamics of timekeeping can be observed in a broad class of operating regimes, making them universal. In this way, our results support the idea that entropy dissipation is not just a prerequisite for measuring time’s passage, but that the entropy dissipated by any clock is quantitatively related to the fundamental limit on that clock's performance. 

\section{Theory: The thermodynamic cost of timekeeping}
\label{sec:theory}

We define the accuracy of a clock as~\cite{Erksi}:
 \begin{equation}\label{eq:acc}
     N := \left( \frac{\ttick}{\Delta \ttick }\right)^{2},
 \end{equation}
where $\ttick$ is the mean interval between successive ticks
and $\Delta \ttick$ is the standard deviation of this interval.
Equivalently, $N^{-1}$ is the Allan variance~\cite{Allan1966} when the observation period is equal to $\ttick$. This is a more severe measure of accuracy than the Allan variance of a much larger number of ticks.  
If Markovian stationarity is assumed, i.e. if successive tick intervals are uncorrelated, $N$ is also the number of ticks before the expected accumulated timekeeping error is equal to one tick interval.

Our objective is to test the measured value of $N$, derived by analysing a series of ticks generated by the experiment, against the prediction of models in which the accuracy of the clock appears as a function of the resources used to drive it. This line of inquiry is inspired by \cite{Erksi}, in which the rate of entropy production and accuracy of an autonomous quantum clock are found to be linearly related (assuming weak coupling), i.e.
\begin{align}\label{eq:Nlinear}
N = \frac{\Delta \Stick}{2\kB}
\end{align}
where $\kB$ is Boltzmann's constant and $\Delta \Stick$ is the entropy generated per tick. This entropy arises due to power being dissipated by the clock, from which we understand that greater power dissipation corresponds to greater accuracy.

In similar spirit we have analyzed a classical model of the optomechanical experiment of Fig.~\ref{fig:1}(c).
In this experiment, the accuracy is ultimately limited by the difficulty of precisely identifying zero-crossings in the presence of thermal noise.
Intriguingly, this classical experiment, despite representing a completely different physical system from the quantum clock of~\cite{Erksi}, obeys a similar relationship between accuracy and entropy.
As shown in Appendix~\ref{App:ClassicalModel}, the maximum accuracy that this classical clock can achieve is
\begin{equation}\label{Nentropy}
     \NC = \frac{2\pi^2}{\kB} \frac{\Tc}{\TN}\Delta \Stick
 \end{equation}
where \TN is the noise temperature of the measurement electronics and \Tc is the temperature of the environment, assumed to be colder than the mechanical effective temperature, which in our experiment is controlled by $P_\mathrm{WN}$.
Whereas $N$ is the accuracy calculated from a sequence of ticks experimentally realised by the clock, $\NC$ is a statistical prediction based on the thermodynamic properties of the setup.

In order to compare the values of $N$ obtained from the experiment with the prediction of the model, we must identify the source of entropy $\Delta \Stick$ in our system. We acknowledge there are various types of entropy emerging from the setup, here, we focus on the entropy in the cavity output signal, as it is directly observable in our temporal traces. Additional entropy contributions are of course produced in the instruments used to control the systems (from the tone that drives the readout cavity to the oscilloscope that measures the cavity output signal). We do not focus on this type of entropy, as it depends on the specific implementation and it is not present in autonomous devices. Finally, there is the source of entropy production that comes from the white noise driving. This is the fundamental entropy dissipated per natural temporal event (tick) in our experiment. Here it is important to note that not all of the power injected in the system will be converted into a useful drive signal, just as not all the energy from a hot bath can be converted into work in a heat engine; some will be dissipated in the environment at the expense of entropy production elsewhere. This does not impact our results as long as the power of the white noise signal used to drive the clock is high enough to make the ticks identifiable above the thermal background. Thus, we estimate the relevant entropy $\Delta \Stick$ from the spectral density of the cavity output signal by computing the area of the spectral density peak located at the membrane's resonance frequency.

\begin{figure}
\includegraphics{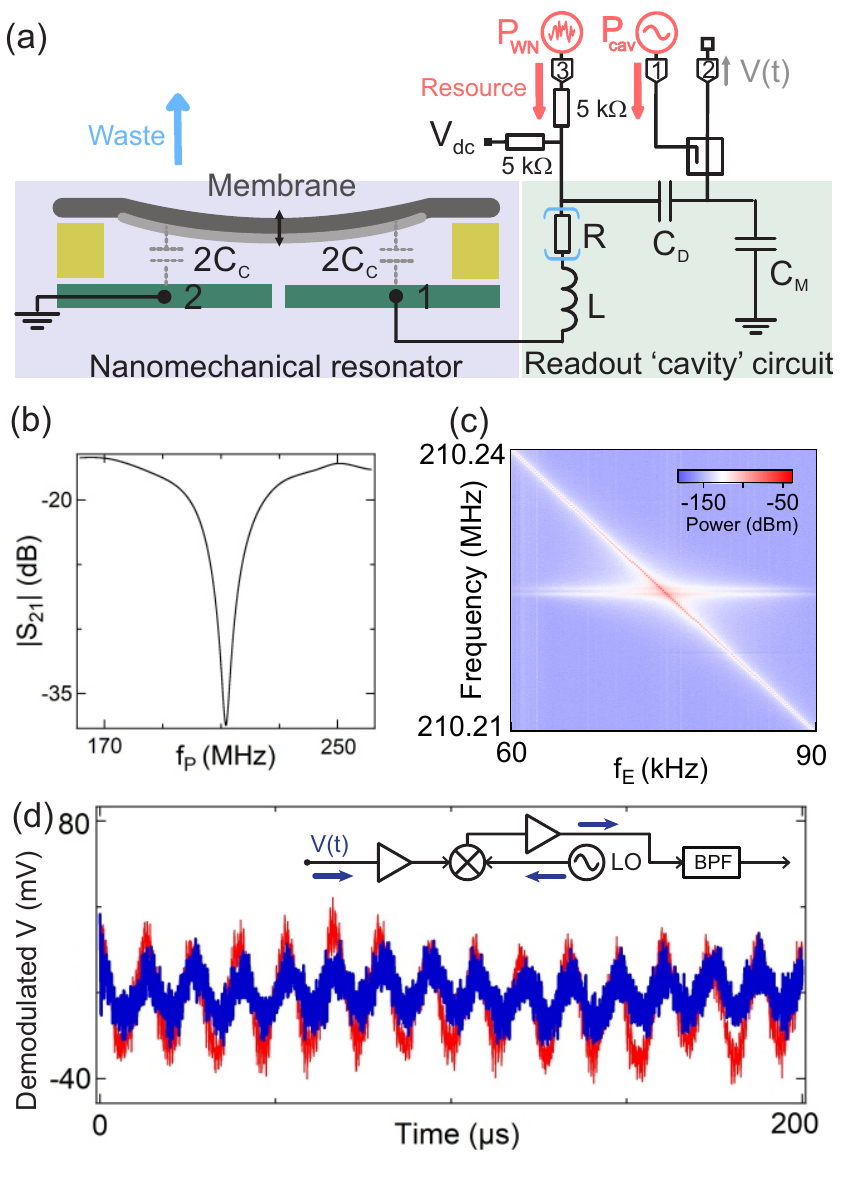}
\caption{\label{fig:2}(a) Experimental setup. A metalized silicon nitride membrane is suspended over two metal electrodes, forming a capacitor C$_C$. One of the electrodes is connected to a RF tank circuit which acts as a readout cavity. Electrode 2 is grounded. The tank circuit is formed from a 223~nH inductor $L$, and two 10~pF capacitors $\CD$ and $\CM$. Parasitic capacitances contribute to $\CM$ and parasitic losses in the circuit are parameterized by an effective resistance R. The cavity can be probed by injecting a RF signal at port 1 via a directional coupler. The output signal is measured at port 2 using a vector network analyser or a spectrum analyser. The membrane's motion can be excited by injecting a signal at port 3. Bias resistors allow a dc voltage V$_\mathrm{dc}$ to be applied to electrode 1. Red (blue) arrows indicate resources (waste) for our system. (b) $|$S$_{21}|$ as a function of probe frequency f$_\mathrm{p}$. (c) One of the mechanical sidebands observed in the spectrum of the cavity output signal when an excitation tone at frequency f$_\mathrm{E}$ is injected at port 3 and swept in frequency whilst the cavity is driven at its resonant frequency via port 1. The sideband power grows when f$_\mathrm{E}$ coincides with the resonance frequency of the membrane f$_\mathrm{0}$. (d) Demodulated readout signal $V(t)$, as a function of time, for $P_\mathrm{cav}=14$~dBm. $P_\mathrm{WN}=0.25$~W and $P_\mathrm{WN}=0.063$~W for the red and blue traces, respectively. The inset shows the demodulation circuit. (LO: local oscillator; BPF: band pass filter)}
\end{figure}

\section{Experimental Setup} 

The vibrating membrane is measured using the setup shown in Fig.~\ref{fig:2}(a).
The membrane, which consists of 50~nm thick SiN metallized with Al, is suspended over two Cr/Au electrodes patterned on a silicon chip, forming a capacitor.
A dc voltage $V_\mathrm{dc}=15$~V is applied to electrode 1, with electrode 2 grounded. Electrode 1 is connected to a RF cavity, which is realised with an inductor and capacitors (Fig.~\ref{fig:2}(a)). As the membrane vibrates, the capacitance $\CC$ between the membrane and the electrodes changes. Driving the RF cavity with a resonant tone, we can probe the membrane's motion by monitoring the cavity's output signal~\cite{Pearson2020}. The cavity is driven by injecting a RF signal via port 1 via a directional coupler. A signal to excite the membrane's motion is incorporated in the circuit via port 3. The experiment is carried out at room temperature at approximately $5 \times 10^{-6}$~mbar.

To determine the cavity's resonant frequency, we measure the scattering parameter $|\mathrm{S}_{21}|$, which is proportional to the reflection from the cavity, as we sweep the frequency of a probe tone $\fP$. The cavity resonance is evident as a minimum in $|\mathrm{S}_{21}|$ (Fig.~\ref{fig:2}(b)).
To identify the mechanical resonance, we perform two-tone spectroscopy.
While driving the cavity at its resonance frequency (i.e. with $\fP=210.3$~MHz) through port 1, we applied another tone of  frequency $\fE$ through port 3 in order to excite the membrane.
The power spectrum of the reflected signal is shown in Fig.~\ref{fig:2}(c) as a function of $\fE$.
The mechanical response is evident as a strong increase in the sideband power at $\fP \pm \fE$ when $\fE$ matches the mechanical frequency $f_0\approx74.5$~kHz~\cite{Pearson2020}.

In order to use the membrane as a thermal clock, we drive the membrane's motion (of the fundamental mode) stochastically by applying a white noise signal of power $P_{\mathrm{WN}}$ and bandwidth 500 kHz through port 3. This white-noise signal is the clock's heating resource. To register the ticks, we must illuminate the cavity, and to do this the resource is a resonant drive tone injected through port 1 with power $P_{\mathrm{cav}}$. We measure the displacement of the membrane in real time by demodulating the cavity output signal $V(t)$. The demodulated signal is measured with an oscilloscope. We show $V(t)$ after demodulation and amplification for two different values of $P_{\mathrm{WN}}$ in Fig.~\ref{fig:2}(d). From these time traces, the ticks of the clock can be identified, and an accuracy can be computed for different values of $P_{\mathrm{WN}}$. 
 
Studying clock performance in the absolute sense is not strictly possible in our system, since this would require us to synchronise multiple clocks (e.g. via the alternating ticks game~\cite{Erkerthesis,Renatello}). We have thus chosen a reference clock that is orders of magnitude faster than the system under investigation in order to resolve the temporal dynamics. In our case, the membrane's frequency is in the kHz regime while our reference clock, the clock of the oscilloscope, operates at a frequency several orders of magnitude higher. Our system constitutes a quasi-autonomous clock, since just with a driving tone, it is able to convert the power of the white noise driving the membrane's motion into the observable ticks of a clock.

\begin{figure*}
\includegraphics[scale=1]{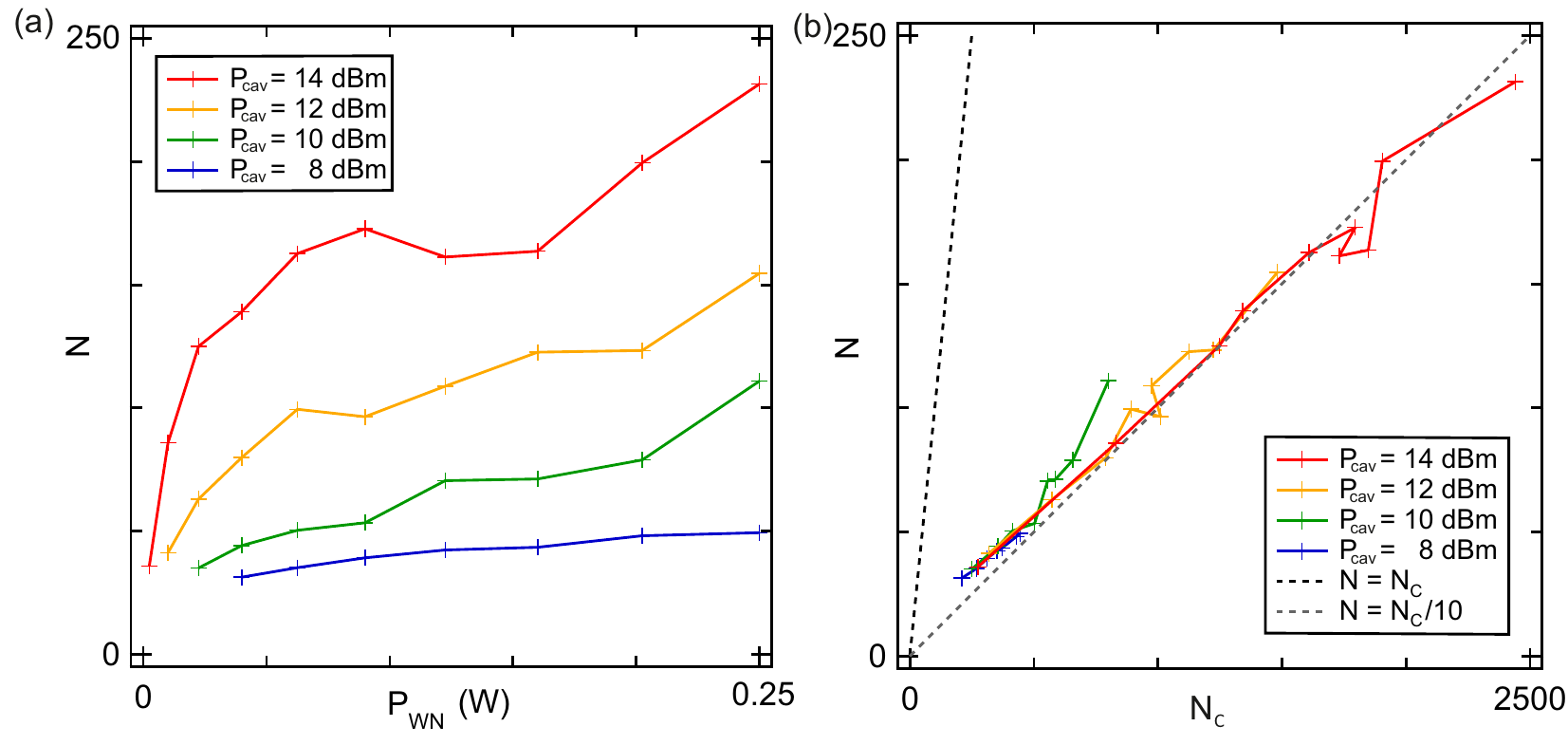}
\caption{\label{fig:acc1} (a) Accuracy $N$ vs white noise power $P_\mathrm{WN}$ for $P_\mathrm{cav}$ in the range $8$ to $14$~dBm.  (b) Accuracy $N$ vs the accuracy predicted from the classical model $\NC$ for $P_\mathrm{cav}$ in the range 8 to 14~dBm. The black dotted line is a guide to the eye showing the slope corresponding to $N=\NC$. The grey dotted line is chosen to approximate the slope of the displayed curves.}
\end{figure*}

\section{\label{Results} Results}

Ticks are generated from time records of the demodulated voltage signal as shown in Fig.~\ref{fig:2}.
Each tick corresponds to an upward zero-crossing of this signal.
In principle, these zero-crossings could be identified in nearly real time using a threshold detector with an appropriate input filter.
In practice, we acquired the entire voltage record and identified ticks in post-processing, in order to be able to study the effects of different filter and threshold settings.

At each setting of $P_\mathrm{cav}$ and $P_\mathrm{WN}$, a record of raw data with a duration of 1~s was stored.
In order to suppress noise, each record was then digitally filtered using a band-pass filter of 75~kHz bandwidth centred at $f_0$.
This bandwidth, which is nearly equal to $f_0$, is sharp enough to remove much of the electronic noise, and thus avoids triggering false upward zero-crossings, but has a fast enough ringdown to ensure that successive ticks are nearly independent.
In a real-time clock, it could be implemented using an analogue filter.
To extract $N$ for each record, the upward zero-crossings were identified in order to generate a sequence of tick intervals, and the resulting standard deviation $\Delta \ttick$ was substituted into Eq.~\eqref{eq:acc}.


The results of this analysis are shown in Fig. \ref{fig:acc1}(a) as a function of $P_\mathrm{cav}$ and $P_\mathrm{WN}$.
For small values of $P_{\mathrm{WN}}$, we see that $N$ increases approximately linearly with $P_{\mathrm{WN}}$. This can be understood intuitively: a stronger drive makes the mechanical oscillations easier to distinguish from the noise. As $P_{\mathrm{WN}}$ increases further, the linear relationship breaks down and the accuracy shows signs of saturating. This is to be expected due to noise in the circuit leaking from the heating tone and the membrane's motion entering the non-linear regime, effects which do not allow a continued increase of $N$.

As $P_\mathrm{cav}$ increases, the linear increase of $N$ as a function of $P_{\mathrm{WN}}$ shows a larger gradient. This is because an increased $P_\mathrm{cav}$ enhances readout. Above $P_\mathrm{cav}=14$~dBm, however, demodulated $V(t)$ shows significant fluctuations, leading to the saturation of $N$ at smaller values of $P_{\mathrm{WN}}$ (see Appendix \ref{highdbm}). The time traces corresponding to $P_\mathrm{cav}<8$~dBm are too noisy for ticks to be identified (see Appendix \ref{lowdbm}). The oscilloscope's sampling rate was 40~MSa/s, giving a resolution of 25~ns to the acquired time traces. Given the frequency of the membrane, this resolution sets an upper limit to the measurable accuracy of $N \lesssim 290,000$; however, as seen from Fig.~\ref{fig:acc1}(a), experimental values of $N$ are less than a hundredth of this limit.

To test the predictions of the classical clock model, we now compare the measured $N$ with the predicted accuracy $\NC$ according to  Eq.~\eqref{Nentropy}.
The relevant entropy arises from the electrical power dissipated in the amplifier circuit by the optomechanical sidebands that contain the displacement information.
As shown in Appendix~\ref{App:ClassicalModel}, the ratio $\Delta \Stick/\TN$ can be calculated from the same demodulated voltage record used to identify ticks.
To do this, each record is first numerically transformed to generate a power spectrum.
The entropy $\Delta \Stick$ is then calculated from the integrated power within a 10~kHz window centered on the signal frequency $\ff$; the noise temperature $\TN$ is calculated from the average spectral density well away from this frequency (see Eq.~\eqref{eq:SNspectrum}). The physical temperature of the measurement circuit is taken as $\Tc=300$~K.

We have compared the values obtained for $\NC$ with the accuracy $N$ computed as in Eq.~\eqref{eq:acc} (Fig.~\ref{fig:acc1}(b)). Our results confirm that increasing accuracy require increasing $\Delta \Stick$, and show the linear relation predicted by Eq.~\eqref{Nentropy}. However, the constant of proportionality, for all heating and illumination powers shown here, is approximately ten times smaller than predicted. Since Eq.~\eqref{Nentropy} represents an upper bound on the clock's efficiency, this discrepancy is not inconsistent with the theory. It probably indicates that identifying the zero-crossings, which does not use all the information in the voltage record, is not an optimal procedure for identifying ticks.

\section{\label{discussion} Discussion}

Our experiment is simple enough to account for the thermodynamic resources used, like in Ref.~\cite{Brunelli2018}, and at the same time our system is too complex to be modelled by a simple open quantum systems approach. 

The results in Fig.~\ref{fig:acc1} showcase an important  relation between the accuracy and the entropy production that should be present in the most fundamental clocks \cite{Erksi}, both in a quantum and a classical model. The accuracy is only a lower bound on the entropy creation, making it entirely possible for the system to dissipate more entropy at higher drive powers without providing more accurate ticks. The fact that we nonetheless see such a consistent linear relation between the accuracy and the entropy production for a considerable range of cavity and white noise drives, indicates that our clock's performance is close to optimal and that we are correctly identifying the relevant entropy contributions.

Our clock provides a steady stream of ticks that are identified from cumulative events; it would defeat the purpose of a clock if only a finished sequence of events can be used retroactively for the identification of ticks. That would rather correspond to the concept of a stopwatch, where upon interrogation one obtains a good estimate of how much time has elapsed between initialisation and interrogation, but does not provide a continuous temporal reference frame. Although the system is not fully autonomous, because a cavity drive is necessary for readout, it presents a perfect testbed for generating a stable time-ordered signal by exploiting thermal non-equilibrium. In fact, any system that acts as a register is expected to consume work, as it would inevitably require to perform measurements of irreversible events ~\cite{Guryanova2020idealprojective}.



Any thermally irreversible process could be used as a clock \cite{milburn2017quantum}, e.g.\ simply by observing the progress of equilibration as a proxy for time. We propose that an operational definition for a good clock is a system that reduces the linear slope of the accuracy-dissipation relation and keeps it linear for accuracy as high as possible. This is consistent  with another recent finding Ref.~\cite{upcomingManu}, which shows that clockwork complexity can be used to decrease that linear slope and to increase the saturation point, beyond which extra dissipation will not correspond to a better clock quality.

We should also note an interesting relation to the phenomenon of stochastic resonance \cite{RevModPhys.70.223}, where noise can push a signal beyond a detection threshold and in this way increase the signal quality. Superficially our experiment presents a similar scenario, since we inject noise to create a periodic signal in time. The main difference, however, is that we do not add noise to the output signal with a constant read-out limitation, but rather the opposite: we feed our noise directly into the physical system producing the output signal and modifying the read-out mechanism to optimally reveal it in a noisy background. Nonetheless, it will be interesting to see if these techniques can be fruitfully adapted to our setup.

The observed relationship between drive power and accuracy (Fig.~\ref{fig:acc1}) is in qualitative agreement with the relation stemming from the oversimplified model in Ref.~\cite{Erksi}, and with the prediction of our classical model. Our results also corroborate the notion that the quality of the arrow of time is indeed limited by the entropy dissipated by a clock. As described in Ref.~\cite{Erksi}, the linear relation between accuracy and entropy production tends to break down at some point. We have observed this effect in our experiment, most likely due to the membrane's motion entering the non-linear regime at high drive powers or due to other non-linearities playing a more significant role in the circuit. Below that threshold, our observed relationship between drive power and accuracy points towards a universal relation, in both quantum and classical regimes, between entropy production and clock accuracy. 

\section{\label{conclusions} Conclusion and Outlook}

In this work, we have demonstrated a thermomechanical clock which allowed us to reveal a universal relation in the thermodynamics of timekeeping. We have first showed that the heating resource introduced to drive the clockwork of our optomechanical setup enhances the accuracy of the clock signal. Modelling our system classically, we have then found that the linear relationship between clock accuracy and entropy production, derived in an idealised quantum setting, is found to hold in the classical regime. The universality of this relation provides a clear link between the entropy dissipated by the clock and the quality of the arrow of time.

As an exciting avenue for future investigation, one can imagine interpreting the system as a heat engine, instead of a clock. Since the oscillations of the membrane can induce a current, they are able to produce work, thus mimicking a heat engine that converts unstructured noise into regular beats. For a system of this scale, work fluctuations become crucial, in contrast to a classical macroscopic engine, for which the power delivered in each stroke is approximately the same. This opens up the opportunity of studying work fluctuation relations as well as deriving rates for heat to work conversion. Finally, it would be interesting to see if the noise (heat) driving the membrane could be harnessed from the environment, rather than being input from a characterised source. In this way one would be able to say that the system is truly performing as a useful engine.

\begin{acknowledgments}

We acknowledge useful discussions with G.~Milburn, M.~Lock, and J.~Parrondo, and F. Vigneau's contribution to the experiment. This work was supported by the Royal Society, EPSRC Platform Grant (EP/R029229/1), the ERC (818751) and FQXi Grant (FQXi-IAF19-01). This publication was also made possible through support from Templeton World Charity Foundation and John Templeton Foundation. The opinions expressed in this publication are those of the authors and do not necessarily reflect the views of the Templeton Foundations. MH acknowledges funding from the Austrian Science Fund (FWF) through the START project Y879-N27, FQXi Grant number FQXi-IAF19-03-S2 and the ESQ  Discovery Grant \emph{Emergent time - operationalism, quantum clocks and thermodynamics} of  the  Austrian  Academy of  Sciences  ({\"O}AW). P.E. and Y.G. acknowledge funding from the Austrian Science Fund (FWF) through the Zukunftskolleg ZK03.

\end{acknowledgments}

\vfill

\providecommand{\noopsort}[1]{}\providecommand{\singleletter}[1]{#1}%

\newpage
\appendix
\onecolumngrid
\newpage

\section{Electromechanical system}

The silicon nitride membrane is 50~nm thick and has an area of 1.5~mm $\times$ 1.5~mm. $90\%$ of the area of the membrane is metalized with 20~nm of Al.  We suspend this membrane over two Cr/Au electrodes patterned on a silicon chip. The capacitor formed between the electrodes and the metalized membrane, which depends on the membrane's displacement, leads to coupling between the cavity and the mechanical motion. The RF circuit is modelled and characterised in Ref.~\cite{Pearson2020}. The entire setup forms a three-terminal circuit with input ports 1 and 3 and output port 2. We used a vector network analyzer to measure the scattering parameter (Fig.1b), a spectrum analyzer to measure power spectra (Fig.1c) and an oscilloscope to measure the displacement as a function of time (Fig.1d).

\section{Entropy-accuracy relation for a thermomechanical clock}\label{App:ClassicalModel}

This Appendix derives the entropy-accuracy relation, Eq.~\eqref{Nentropy}, which is tested in the main text. We do this by considering two classical clock models.
The first is a very simple clock that uses the filtered Johnson noise of a hot resistor.
The second is the optomechanical clock --- an elaboration of the Johnson noise clock which is realised in our experimental setup.
As shown below, both designs obey the same relation, which in turn resembles previously derived relations for classical~\cite{Barato2015} and quantum~\cite{Erksi} clocks.

In both models, the clock must derive ticks from a periodic but noisy voltage record.
We ask the question: how precisely can any clock identify a tick instant from a segment of this record?
From the perspective of the clock, this is clearly a problem of phase estimation.
From the $n$th segment of the record, an error $\delta \phi_n$ in estimating the phase leads to an error $\delta t_n = \ttick \delta \phi_n / 2 \pi$ in identifying the corresponding tick instant $t_n$.
Thus from Eq.~\eqref{eq:acc}, the clock accuracy in any classical model is related to the phase error by
\begin{align}
\NM	
	&:=	\frac{4\pi^2}{\langle (\delta \phi_n)^2 \rangle}.	\label{eq:N2}
\end{align}
since the tick uncertainty $\Delta t$ is by definition the root-mean-square value of $\delta t_n$.
Furthermore, we require that successive ticks be statistically independent, which means that every tick must be derived from a non-overlapping segment of the record.
In what follows, we construct models for $\delta \phi_n$ for two physical scenarios and thus estimate the accuracy of those clock models. 

\subsection{Measuring time from filtered Johnson noise \label{Sec:JohnsonModel}}

\begin{figure*}
	\includegraphics{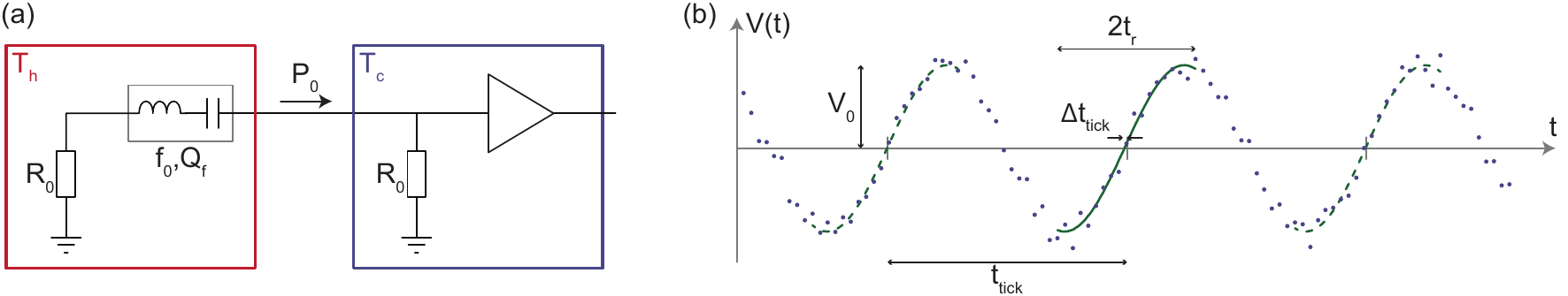}
	\caption{\footnotesize{Simple electrothermal clock. (a) The setup. The Johnson noise of resistor $R_0$, in equilibrium with a hot bath at temperature~$\Th$, is filtered to pass frequency $\ff$ with bandwidth $\ff/\Qf$. The resulting signal, whose power is $P_0$, is passed to a matched resistor and amplifier at temperature $\Tc$.
	(b) From the noisy voltage record (points) seen by the amplifier, we can generate clock ticks by estimating the zero crossing of each cycle using a sinusoidal fit (lines). Here $\Delta t$ marks the sampling interval, $\ttick=1/\ff$ is the average tick interval, $\pm \trr$ is the fit range, and $\Delta \ttick$ is the fit uncertainty.}
	\label{FigS:ThermoClock}
	}
\end{figure*}

Figure \ref{FigS:ThermoClock} shows a design for a thermodynamical clock based on Johnson noise.
The clock contains two heat baths at temperatures $\Th$ and $\Tc$. 
Inside the hot bath, at temperature $\Th$, is a resistor $\RR$, which is connected via a matched transmission line to an ideal voltage amplifier located in the cold bath at temperature $\Tc$.
To ensure an impedance match and thus prevent reflections from the end of the transmission line, an equal resistor \RR is connected to the amplifier input. A reflective band-pass filter is placed in the transmission line, centered at frequency $f_0$ and with quality factor $\Qf$, so that it passes frequencies in a bandwidth of $f_0/\Qf$ near the center frequency. The combined Johnson noise of the two resistors leads to an incoherent voltage oscillation at the cold amplifier input, whose peak amplitude $\VS$ satisfies
\begin{equation}
\langle \VS^2 \rangle	= 2\kB (\Th+\Tc) \RR \, \frac{\ff}{\Qf}.
\label{eq:VS}
\end{equation}
where $\langle \cdot \rangle$ denotes an average over many oscillations.
Each oscillation cycle corresponds to one tick of the clock.
Demarcating each cycle accurately requires a large oscillation amplitude, meaning that a larger power is dissipated in the cold resistor; this is the thermodynamic price that we aim to quantify.

The amplifier measures the input voltage $V(t)$ as a function of time $t$ (Fig.~\ref{FigS:ThermoClock}(b)).
To generate a timing signal, the clock's task is to identify ticks from particular instances of the record, for example, those instants at which the upward crossings of the $t$-axis occur.
This is the phase estimation problem described above.
The reason that a perfect estimate is impossible even in principle is that the record is contaminated by voltage noise, including the broadband Johnson noise of the cold resistor.

How should the clock best perform a phase estimate, given a segment from the noisy voltage record?
The answer is to perform a maximum-likelihood estimation.
If the noise is uncorrelated and has a Gaussian distribution, as expected for broadband Johnson noise, this means a least-squares fit to the data~\cite{Press2007}.
No implementation of the clock can perform better than this.

To this end, we imagine that we have obtained some experimental data; we discretize the time interval in the record into pieces around the expected tick locations (the upward crossings), and fit one curve for each tick of the clock, such that for $n$ ticks we fit $n$ curves. 
For a particular tick we imagine fitting the function 
\begin{align} \label{eq:ffit}
V(t|\phi_n) = V_0 \sin (2\pi f t + \phi_n)
\qquad\;\;
\begin{matrix}
2n\cdot\ttick - \trr \le t \le 2n\cdot\ttick + \trr
\qquad\text{for}
\qquad n\in \mathbb{Z}
\end{matrix}
\end{align}
where $V_0$ is the oscillation amplitude; $f$ is the frequency; $\phi_n$ is the phase; and where we have chosen to fit the $n$-th tick to a function over the interval $2\trr$ (see Fig.~\ref{FigS:ThermoClock}(b)). 
The parameters $V_0$ and $f$ can be estimated over several recent oscillation cycles because they are slowly varying properties and are therefore not determined by the noise over a single cycle.
The only parameter to fit is thus the phase $\phi_n$, which motivates the notation $V(t|\phi_n)$ as per the prescription in ~\cite{Press2007}. For a particular data set $\mathcal{D}$, the optimal value of the parameter for the $n$-th tick, denoted $\phi_n^*$, is the one that minimises the $\chi^2$ function, defined as
\begin{align}
    \chi^2(\phi_n) = \sum_{i}^{} 
    \Bigg[\frac{V_i - V(t_i|\phi_n)}
    {\sigma_i}
    \Bigg]^2
    \,,\label{eq:chi}
\end{align}
where $i$ labels the data points and ranges over the total number of data-points, and $\sigma_i$ is the vertical standard deviation of each point.
The uncertainty is then determined by $\Delta \chi ^2 = 1$ and the curvature parameter $\alpha$, and follows the expression:
\begin{equation}
\Delta \phi:=
\sqrt{\langle (\delta \phi_n)^2 \rangle} :=
\sqrt{\Delta\chi^2}\,
\alpha^{-1/2}.
\label{eq:phi}
\end{equation}
The curvature parameter is calculated from the fitted function and the experimental points $i$ as
\begin{equation}
\alpha := \sum_i \frac{1}{\sigma_i^2} \left(\frac{\partial V(t_i|\phi_n)}{\partial \phi_n}\right)^2.
\label{eq:alpha}
\end{equation}
A final value for Eq.~\eqref{eq:phi} would be obtained by evaluating $\alpha$ at the fitted parameter $\phi_n^*$ which minimises Eq.~\eqref{eq:chi} and choosing $\Delta \chi^2$ such that it corresponds to the desired confidence interval. Since we are in the business of constructing a model for the accuracy (i.e.~we are not analysing the fit of a particular data set), we must make a statement that is reasonable for all data sets $\{\mathcal{D}\}$ that may emerge from this setup. To do this, we must make a few additional assumptions.
First, we are interested in a situation where the oscillation frequency is sharply defined, i.e. $\Qf \gg \langle V_0^2 \rangle / \sigma_i^2$, which means that within a single cycle $\sigma_i$ is dominated by the broadband noise at the amplifier input and therefore takes a constant value $\sigma$ for all data points.
Next, we imagine that the $n$ ticks are fitted by choosing $n$ windows (or regions) of length $2\trr$ where $\trr = 1/2f_0$, and the $\chi^2$ minimisation gives us the value of the crossing $\phi_n^*$ for each tick. 
To calculate $\alpha$ in any such region, we idealise Eq.~\eqref{eq:alpha} by imagining a continuum of data points, and thus convert the sum to an integral normalised by $\Delta t$, the sampling interval. 
This gives 
\begin{align}
\alpha 	&=	\frac{1}{\sigma^2 \Delta t}
\int_{-\trr}^{\trr} 
\left(
\frac{\partial V(t|\phi_n)}{\partial \phi_n}
\right)^2 dt
\label{eq:step1int}
\\
&=	\frac{V_0^2}{\sigma^2 \Delta t}
\int_{-\trr}^{\trr} \cos^2 (2\pi f_0 t+ \phi_n) \;dt\
		\\
	&=\frac{V_0^2}{\sigma^2 \Delta t}
 \left(
\trr+
\frac{\cos (2\phi_n)\sin(4\pi f_0\trr)}
{4\pi f} 
\right)
\\
& = \frac{V_0^2}{2\sigma^2 \Delta t f_0}
\end{align}
where the last step assumes $\trr$ has been chosen at the optimal value of $1/2f_0$, and without loss of generality the zero of $t$ has been chosen at the centre of the fit interval.

Notice that choosing to fit the function in windows of width $2\trr =1/f_0$ has resulted in an expression for $\alpha$ that is independent of the fitted parameter $\phi_n$. Indeed, the integral in Eq. \eqref{eq:step1int} is independent of $\phi_n$ for all integration regions of width $2\trr = 1/f_0$, regardless of where they are centred. Thus, knowledge of the membrane frequency $f_0$ implies that the standard error in the fitted parameter $\phi_n$ is  only related to the physical parameters set for the experiment. Also note that on converting the sum to an integral, we would expect the expressions to be approximately equal. Observe that the right-Riemann sum $\alpha \Delta t $ would overestimate the integral of any monotonically increasing function in the interval, while underestimating for a monotonically decreasing function. If the parameter is fitted such that it falls roughly within the centre of the window each time (i.e.~we place the window roughly where we expect the crossing), the effects of over and underestimating the symmetric function under the integral will roughly balance out.

To obtain the standard deviation $\Delta \phi_n$, we should take $\Delta \chi^2=1$ in Eq.~\eqref{eq:phi}, giving
\begin{align}
\Delta \phi =
\alpha^{-1/2}.
\label{eq:phi1}
\end{align}
With this, the accuracy in the Johnson-noise model is
\begin{align}
    \NJ &= {4\pi^2}\alpha \\
    &=\frac{2\pi^2 
V_0^2}{\sigma^2 \Delta t f_0} \label{eq:jmodel}
.
\end{align}
The per-point standard deviation depends on the measurement bandwidth of the amplifier and on the system noise.
In the best case, it will be set by the Johnson noise of the cold resistor~\cite{Clerk2010}, giving
\begin{equation}
\sigma = \sqrt{4 \kB \Tc \RR B}
\label{eq:sigma}
\end{equation}
where $B$ is the measurement bandwidth (defined using the single-sided frequency convention), and the factor 4 appears because the bandpass filter presents on open load except near resonance.
In order that successive points are independent but no data is lost, the sampling interval should be related to the bandwidth by $B=1/2\Delta t$.
Thus
\begin{equation}
\sigma^2 \Delta t = 2 \kB \Tc \RR.
\label{eq:sigmaDeltat}
\end{equation}
Substituting into Eq. \eqref{eq:phi1} gives for the phase uncertainty in the interval which we chose to fit
\begin{equation}
\delta \phi_n = \frac{\sqrt{4 \ff \kB \Tc \RR}}{V_0}.
\label{eq:phi2}
\end{equation}
Over many oscillations, $V_0$ fluctuates, but its root-mean-square value is $\VS$, given by Eq.~\eqref{eq:VS}. Substituting this and \eqref{eq:sigmaDeltat} into Eq. \eqref{eq:jmodel} gives us a model for the accuracy of the clock
	\begin{align}
\NJ	&=  \frac{2\pi^2}{\ff\kB \Tc R_0} \langle V_S^2 \rangle\label{eq:N3}\\
	&=	\frac{2\pi^2}{\Qf} \frac{\Th+\Tc}{\Tc}.		        \label{eq:N4}
\end{align}
The clock creates entropy because the power carried by the electrical oscillation is converted to heat in the cold resistor.
The entropy creation rate can be written
\begin{align}
\dot{S}	
		&= \kB\frac{\Th-\Tc}{\Tc}\frac{\ff}{\Qf},
		\label{eq:Sdot}
\end{align}
since the net power transferred is $P_0=\kB(\Th-\Tc)\ff/\Qf$.
Combining this expression with Eq.~\eqref{eq:N4} gives the accuracy in terms of the entropy created:
\begin{align}
\hspace{3cm}
\NJ	&= 2\pi^2 \,\frac{\Th+\Tc}{\Th-\Tc}\, \frac{\Delta S_\mathrm{tick}}{\kB} \\
	&\approx 2\pi^2 \,\frac{\Delta S_\mathrm{tick}}{\kB}		
 \label{eq:NS}\qquad\qquad\qquad 	\text{for $\Th \gg \Tc$.}
\end{align}
where $\Delta\Stick \equiv \dot{S}/\ff$ is the entropy generated per tick. 
This best case scenario (i.e. smallest $\sigma$) provides an upper bound for the best achievable accuracy of an experiment of this type. Thus, we can expect this model to overestimate the accuracy compared to that coming from a live experiment.  
Similar expressions to Eq.~\eqref{eq:NS} hold for a classical clock defined by transitions on a network~\cite{Barato2015} and for an autonomous quantum clock~\cite{Erksi}; however, in both these cases the factor $2\pi^2$ is replaced by $1/2$.


\subsection{Measuring time from an optomechanical signal}

\begin{figure*}
	\includegraphics{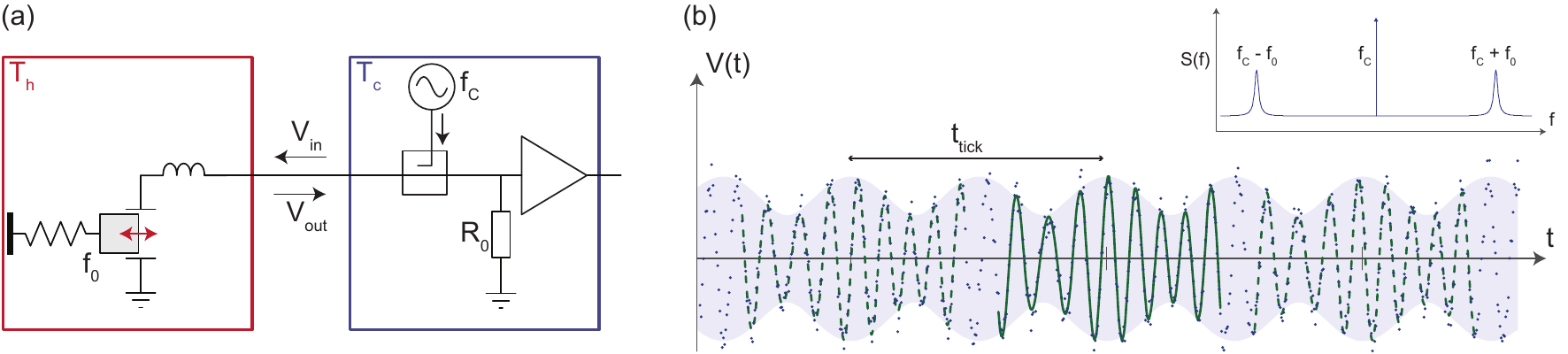}
	\caption{\footnotesize{The thermomechanical clock. (a) The setup. An optomechanical circuit consists of an $LC$ tank circuit whose capacitance, and therefore frequency, is modulated by a thermomechanical resonator at temperature $\Th$. To use the vibrations in a clock, the tank circuit is illuminated by a carrier tone $\Vin$ at frequency $\fc$, giving rise to a reflected signal $\Vout$ which is passed to a cold matched resistor and amplifier. The effect of the vibrations is to modulate the phase of $\Vout$.
(b) Sketch of the resulting voltage record at the amplifier input (points), with fits (lines) from which each tick is extracted. The modulation envelope is indicated by the shaded background. Inset: Power spectrum at the amplifier input, showing uniform noise background, central delta-function peak from the carrier, and two thermomechanical sidebands.
In the experiment, a demodulation circuit was applied after the amplifier (as in Fig.~\ref{fig:2}(d)) because it makes ticks practically easier to identify in the record.
However, the demodulator cannot improve the clock accuracy because it  cannot increase the timing information present in the signal $V(t)$; in fact a detailed calculation would show that the accuracy is unchanged.
For simplicity the demodulator is therefore omitted from this model.
	\label{FigS:OMClock}
	}}
\end{figure*}
In this section we proceed to build a classical model, that predicts the accuracy, which we call $\NC$, for a scheme that is more fitting to our experimental setup. Figure~\ref{FigS:OMClock} shows the optomechanical setup, which serves as the clock of our experiment. 
The clock works by illuminating a tank circuit containing a vibrating membrane with an RF tone of power $\Pc$ (Fig.~\ref{FigS:OMClock}(a)).
The thermal motion of the membrane modulates the phase of the reflected signal, and from this signal the ticks are derived.
This is the principle of the clock realised in our experiment.
The advantage of this clock over the version of Fig.~\ref{FigS:ThermoClock} is that the reflected signal can be increased by increasing $\Pc$ as well as by heating the membrane more strongly.
As this section will show, this clock obeys a similar entropy-accuracy relation to Eq.~\eqref{eq:NS}.
 The voltage incident on the tank circuit is
\begin{equation}
\Vin(t) = \Vc \cos (2\pi \fc t)
\end{equation}
where $\Vc = \sqrt{2R_0\Pc}$ and $\fc$ are respectively the amplitude and frequency of the illumination signal, and the characteristic impedance of the transmission line is assumed equal to $R_0$.
The reflected amplitude is therefore 
\begin{equation}
\Vout(t) = \Gamma \Vc \cos (2\pi \fc t + \beta x(t))
\label{eq:Vout}
\end{equation}
where $\Gamma$ is the cavity reflection coefficient, $\beta$ is the mechanical coupling strength, and $x(t)$ is the instantaneous membrane displacement.
The phase reference plane is assumed to be chosen so that the phase is zero at the membrane's equilibrium position. The membrane vibrates with a mechanical temperature $\Th$.
If its quality factor is high, the mechanical amplitude $x_0$ and phase $\phi$ are approximately constant over one oscillation cycle, meaning that the displacement is
\begin{equation}
x(t)=x_0 \cos (2\pi\ff t + \phi).
\label{eq:x}
\end{equation}
In this experiment, the electromechanical coupling is weak, meaning that $\beta x_0 \ll 1$.
This means that we can substitute Eq.~\eqref{eq:x} into Eq.~\eqref{eq:Vout} and expand to lowest order in $\beta x_0$, giving
\begin{equation}
\Vout(t)=\Gamma \Vc \left[\cos (2 \pi \fc t) - \beta x_0 \sin (2 \pi \fc t) \cos (2\pi\ff t +\phi) \right].
\label{eq:VoutOM}
\end{equation}
In other words, the reflected signal is modulated at frequency $\ff$ with phase $\phi$, as sketched in Fig.~\ref{FigS:OMClock}(b).
Each full cycle of the modulation is one period of the clock. To generate ticks, the clock must identify a particular point of the modulation cycles, which implies it must precisely identify $\phi$.
As in Section~\ref{Sec:JohnsonModel}, we want to know how accurately this can be done in principle. 
Again, we imagine we have obtained a set of experimental data and wish to know how accurately the $n$-th tick can be identified. We proceed by fitting the function
\begin{equation}
V(t|\phi_n)=A_0\cos (2 \pi \fc t) +A_1 \sin (2 \pi \fc t) \cos (2\pi\ff t +\phi_n).
\label{eq:VoutFit}
\end{equation}
in windows of width $1/f_0$ around the expected tick locations. The parameters $A_0$, $A_1$, $\fc$, and $\ff$ can be extracted over several recent cycles, and are thus known values. Therefore, just as in  Section~\ref{Sec:JohnsonModel} we are performing a one-parameter fit. 

We imagine that for some dataset we minimise Eq. \eqref{eq:chi} for the function in Eq. \eqref{eq:VoutFit}, which gives us the optimal parameter $\phi_n^*$. We now want to know: what is the error in this fit given the optomechanical setup we have described? We follow the recipe give in the previous section and proceed to calculate the curvature parameter of our model
\begin{align}
\alpha 	&=	\frac{1}{\sigma^2 \Delta t}\int_{-\trr}^{\trr} \left(\frac{\partial V(t|\phi_n)}{\partial \phi_n}\right)^2 dt\\
		&=	\frac{A_1^2}{\sigma^2 \Delta t}\int_{-\trr}^{\trr} \sin^2 (2 \pi \fc t) \, \sin^2 (2\pi\ff t +\phi_n) \;dt\\ 
		\begin{split}
		&=	\frac{A_1^2}{4 \sigma^2 \Delta t}
		\int_{-\trr}^{\trr}
		1  - \cos (4\pi\ff t +2\phi_n) - \cos (4 \pi \fc t)\\
		&\hspace{2cm}-\frac{\cos  (4\pi(\fc+\ff) t +2\phi_n) }{2} \\
		&\hspace{2cm}+\frac{\cos  (4\pi(\fc-\ff) t -2\phi_n)}{2}\;dt	\label{eq:alphaIntegralLong}	
		\end{split}
\end{align}
where $\trr=1/2f_0$ is the fit range.
Since the fit window extends over many cycles of the carrier tone, i.e. $\trr \gg 1/f_c$ 
the last three oscillatory terms make a negligible contribution to the integral, leaving
\begin{align}
\alpha 	&=  \frac{A_1^2}{4\sigma^2\Delta t} \int_{-\trr}^{\trr} 1  - \cos (4\pi\ff t +2\phi_n) \; dt \\
        &=  \frac{A_1^2}{2\sigma^2\Delta t} \left(\trr-\frac{\cos 2 \phi_n}{4\pi f_0}\,\sin (4\pi f_0 \trr)\right)\\
        &= \frac{ A_1^2}{4 f_0\sigma^2\Delta t}
\end{align}
Since the tank circuit presents an open electrical impedance except at its resonance frequency, the Johnson noise again obeys Eq.~\eqref{eq:sigmaDeltat}, leading to
\begin{equation}
\alpha = \frac{A_1^2}{8\ff\kB\Tc\RR},
\end{equation}
which implies that
$
\delta \phi_n =  \sqrt{8\ff\kB\Tc\RR} / {A_1}
$
and 
\begin{align}
    \NC &= {4\pi^2}\alpha
    = \frac{\pi^2 \langle A_1^2\rangle}{2 \ff \kB \Tc R_0}.\label{eq:NOM}
\end{align}

To connect this to thermodynamic quantities in the experiment, we recognise that $A_1$ is related to the combined power $\PSB$ in the two sidebands by
\begin{equation}
\PSB=\frac{\langle A_1^2 \rangle}{4\RR}.
\label{eq:PSB}
\end{equation}
Entropy is created because the reflection from the tank circuit containing the hot resonator leads to irreversible heating in the cold resistor.
Equation~\eqref{eq:VoutOM} and the inset of Fig.~\ref{FigS:OMClock}(b) show that there are potentially two contributions to the heat: the reflected carrier, which is a coherent monochromatic tone at frequency $\fc$; and the two incoherent sidebands centred at $\fc \pm \ff$.
However, the carrier contains no information about $x(t)$.
In principle (although this was not implemented in our experiment) a narrowband filter could be used to direct this portion of the spectrum back towards the tank circuit without affecting the accuracy of the clock.
Thus the reflected carrier does not contribute to the fundamental entropy cost of the clock.
Instead, the unavoidable entropy increase is determined by the two sidebands, which dissipate heat $\PSB$ in the cold resistor.
The entropy creation rate is
\begin{equation}
\dot{S}= \frac{\PSB}{\Tc}.
\end{equation}
In contrast to Eq.~\eqref{eq:Sdot}, there is no decrease of entropy in the hot element, because illuminating the membrane at the cavity frequency does not cool it.
Thus we can re-express Eq.~\eqref{eq:NOM} in terms of the entropy generated per tick, leading to
\begin{equation}
\NC = \frac{2\pi^2}{\kB} \Delta \Stick
\label{eq:SNOM}
\end{equation}
where
\begin{equation}
\Delta \Stick = \frac{\PSB}{\ff\Tc}.
\label{eq:SSOM}
\end{equation}
Equation~\eqref{eq:SNOM} is the fundamental entropy-accuracy relation for the optomechanical clock. 

There is one more adjustment which must be made to compare Eqs.~(\ref{eq:SNOM}-\ref{eq:SSOM}) to experiment.
The derivation above assumed that the amplifier noise is much less than the Johnson noise of the cold resistor.
Although this is perfectly possible, it is also common (and is the case in our experiment) that other noise sources contribute, leading to  a decrease in accuracy that reflects technical imperfections in the voltage measurement rather than any fundamental bound.
To account for this possibility, Eq.~\eqref{eq:SNOM} should be generalized to
\begin{equation}
\NC = \frac{2\pi^2}{\kB} \frac{\Tc}{\TN}\Delta \Stick
\label{eq:SNOM2}
\end{equation}
where $\TN$ is the effective temperature, including the Johnson noise of the cold resistor, determined by the noise in the record.

To evaluate Eq.~\eqref{eq:SNOM2} from the experiment, we express its components in terms of the output signal's power spectrum $\mathcal{S}_{VV}$, which is proportional to the modulus squared of the Fourier transform of the record $V(t)$. In this language, the effective temperature is given by
\begin{equation}
\TN = \frac{\mathcal{S}_{VV}^\mathrm{(N)}}{4\kB\RR}.
\end{equation}
Here $\mathcal{S}_{VV}^\mathrm{(N)}$ is the single-sided average spectral density of the noise in the Fourier transformed signal, i.e. the average background level of the power spectrum.
In terms of the power spectrum, the heat $\PSB$ in the cold resistor is given by integrating the excess spectral density (i.e. the signal) above the noise background, the integral running over both sidebands, $\int \SVV^\mathrm{(S)} (f) \, df$. Thus the classical model predicts the accuracy from the experimental data to be 
\begin{align}
    \NC &= \frac{2\pi^2}{\ff} \frac{A_1^2}{\SVVN} \\
    &= \frac{8\pi^2}{\ff} \frac{\PSB R_0}{\SVVN} \\
        &= \frac{8\pi^2}{\ff} \frac{ \int \SVV^\mathrm{(S)} (f) \, df}{\SVVN} \label{eq:SNspectrum}
\end{align}
where $\SVV^\mathrm{(S)}$ is the and the second line follows from Parseval's theorem.
In practice, our analysis applies Eq.~\eqref{eq:SNspectrum} to the record of the demodulated voltage as in Fig.~\ref{fig:2}(d).
Since demodulation does not change the signal-to-noise ratio, Eq.~\eqref{eq:SNspectrum} remains valid, with the integral now taken over the single signal peak.

\begin{figure*}
\includegraphics[scale=1]{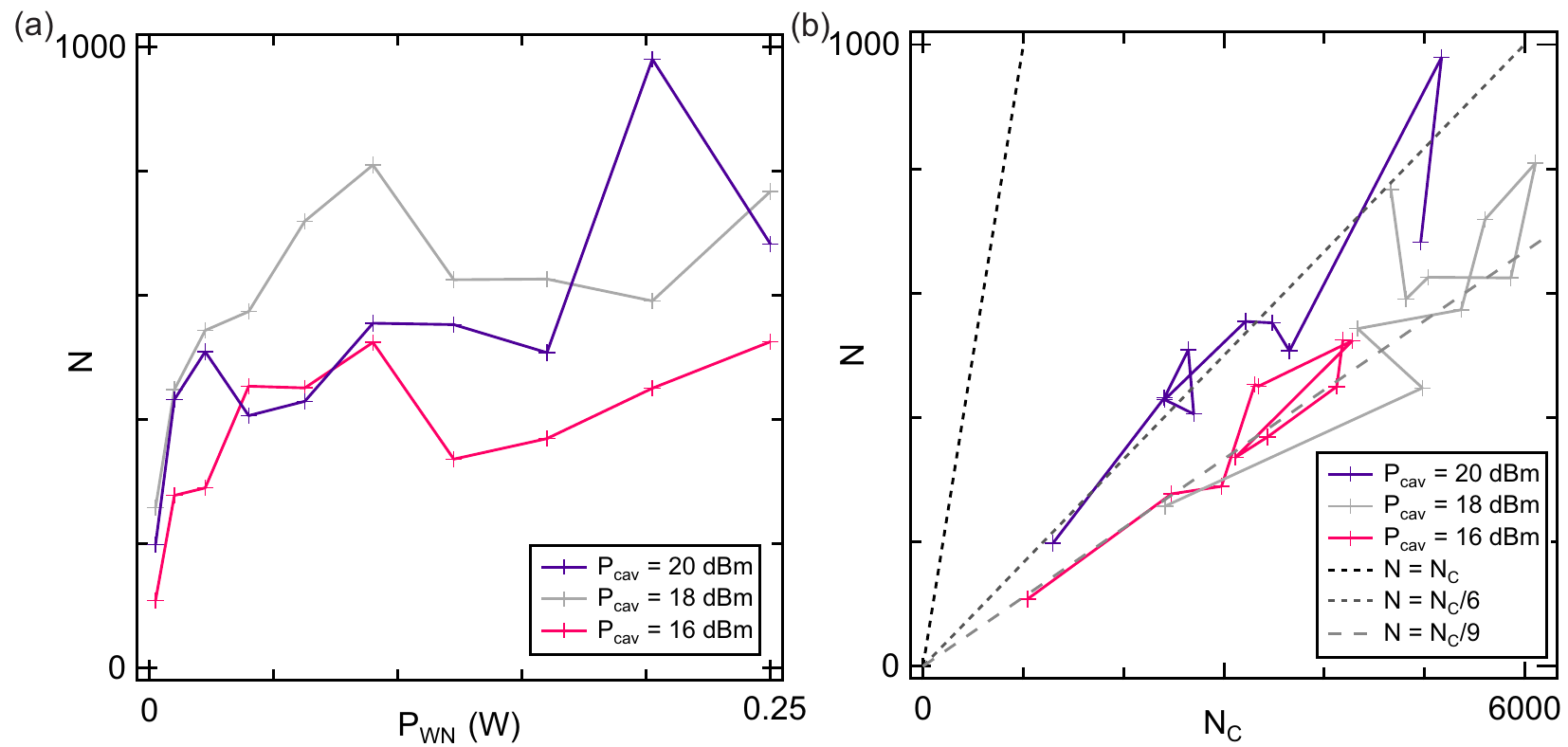}
\caption{\label{fig:4} (a) Accuracy N of the clock vs the white noise power $P_\mathrm{WN}$ for cavity drive powers in the range 16 to 20~dBm.  (b)~Accuracy N of the clock vs N$_\mathrm{C}$ for cavity drive powers in the range 16 to 20~dBm.The black dotted line is a guide for the eye to show the expected gradient should the extracted accuracy and the theoretical accuracy predicted from the entropy production be equal. The dark grey dotted line shows the approximate gradient of the 20~dBm data of N = N$_\mathrm{C}$/6 and the light grey line shows the approximate gradient of the 16 and 18~dBm data of N = N$_\mathrm{C}$/9.} 
\end{figure*}

\section{Overdriving the membrane}
\label{highdbm}
Above 14~dBm we enter the nonlinear regime of the membrane's motion. As can be seen from Fig. \ref{fig:4}(a), for these higher drive powers the relationship between accuracy $N$ and white noise power $P_\mathrm{WN}$ is more erratic. The general trend of accuracy increasing with $P_\mathrm{WN}$ is still there for the lower values of $P_\mathrm{WN}$ but then the accuracy saturates and unstable dynamics dominates the motion of the membrane. As discussed in the main text the saturation of accuracy is to be expected. Surprisingly, Fig. \ref{fig:4}(b) shows that the higher power measurements show better agreement with the theoretical predictions of Eq.~\eqref{Nentropy} (or equally Eq.~\eqref{eq:SNOM2}) with a gradient of 1/9 for the 16 and 18~dBm data sets and 1/6 for the 20~dBm data set.

\section{Measurements at low cavity illumination power}
\label{lowdbm}

\begin{figure}
\includegraphics[scale=1]{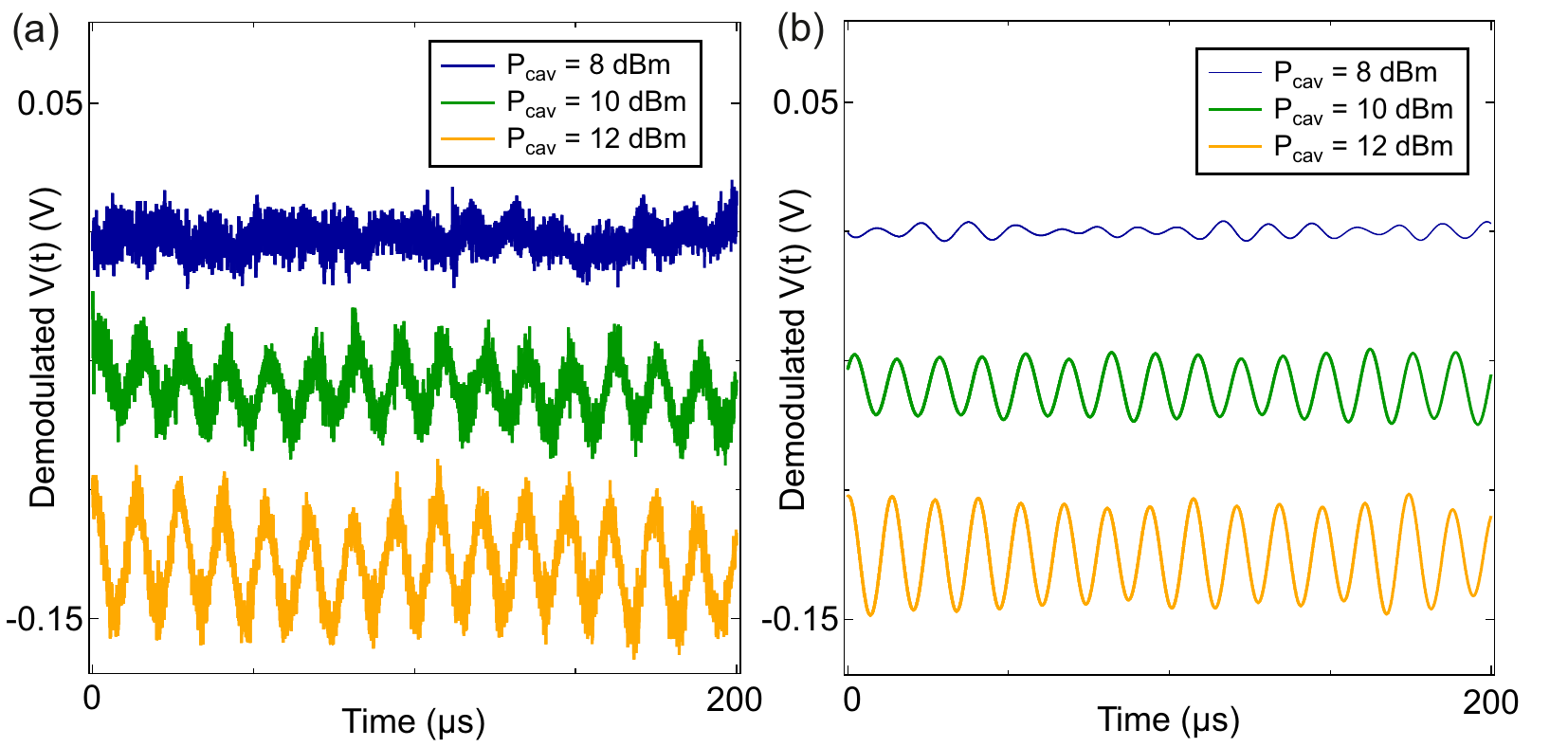}
\caption{\label{fig:5} The demodulated cavity output signal before (a) and after (b) filtering as a function of time for input powers of 8, 10, and 12~dBm to the cavity with 0.25~W white noise input from port 3. The traces are vertically offset for clarity.}
\end{figure}

Figure~\ref{fig:acc1} only shows measurements down to a drive power of $P_\mathrm{cav}=8$~dBm. This is due to the fact that below this drive power the signal is weak and identifying the oscillations becomes difficult. Example traces corresponding to a white noise power of 0.25~W with 8, 10 and 12~dBm drive powers are shown in Fig.~\ref{fig:5} before (a) and after (b) filtering. As can be seen in the signal for a 12 and 10~dBm drive oscillations can easily be identified, however they are much fainter in the 8~dBm signal.


\section{System Noise Temperature}
\label{noisetemp}
\begin{figure*}
\includegraphics[scale=1]{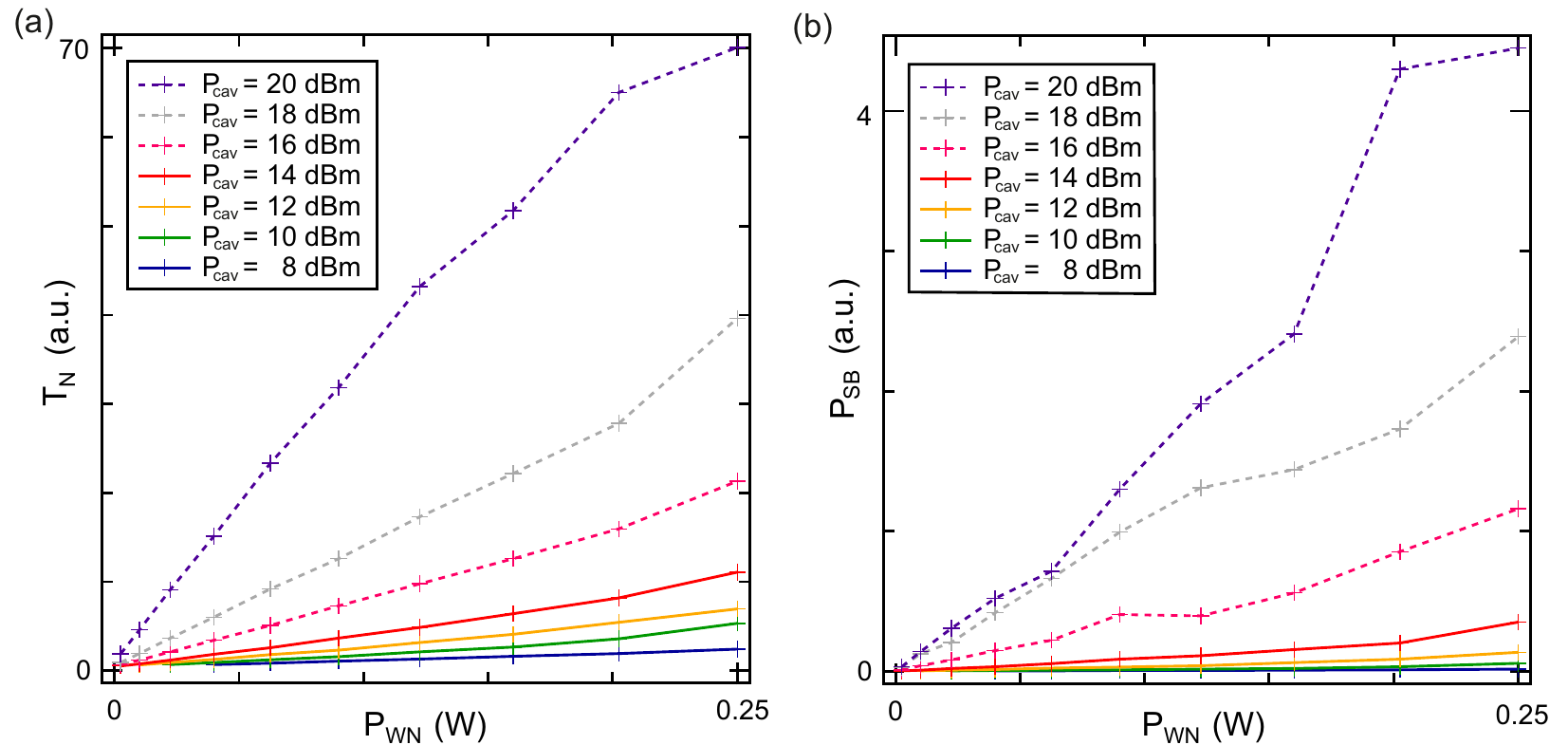}
\caption{\label{fig:6} (a) Noise temperature of the system, $\TN$, increasing as the cavity drive power and white noise power increase for cavity drive powers in the range 8 - 20~dBm. (b)~The increase in the power of the sideband with increase in the white noise power for different cavity drive powers in the range 8 - 20~dBm. For both (a) and (b) the higher drive powers have dotted lines and the data shown in the main text has full lines.}
\end{figure*}

As can be seen from Fig.~\ref{fig:6}(a) the noise temperature of the system increases both for an increased cavity drive or an increased white noise power, $P_\mathrm{WN}$. Fig.~\ref{fig:6}(b) shows that the power in the sideband, used in the entropy calculations (Eq.~\eqref{eq:SNspectrum}), increases approximately linearly with $P_\mathrm{WN}$.   

\end{document}